\documentclass[conference]{IEEEtran}
\IEEEoverridecommandlockouts
\usepackage{cite}
\usepackage{amsmath,amssymb,amsfonts}
\usepackage{algorithmic}
\usepackage{algorithm}
\usepackage{graphicx}
\usepackage{textcomp}
\usepackage{xcolor}
\usepackage{mathrsfs}
\usepackage{hyperref}
\usepackage{booktabs}  %  引入三线表宏包
\def\BibTeX{{\rm B\kern-.05em{\sc i\kern-.025em b}\kern-.08em
    T\kern-.1667em\lower.7ex\hbox{E}\kern-.125emX}}
\begin{document}

\title{Heterogeneity-aware Cross-school Electives Recommendation: a Hybrid Federated Approach}

\author{
\IEEEauthorblockN{
Chengyi Ju,
Jiannong Cao,
Yu Yang, 
Zhen-Qun Yang,
Ho Man Lee}
\IEEEauthorblockA{\textit{Department of Computing}, 
\textit{The Hong Kong Polytechnic University}\\
Hong Kong, China \\
\{juchengyi80, abcdefg61801\}@gmail.com, \{jiannong.cao, cs-yu.yang, zq-cs.yang\}@polyu.edu.hk}
}

\maketitle

\begin{abstract}
In the era of modern education, addressing crossschool learner diversity is crucial, especially in personalized recommender systems for elective course selection. However, privacy concerns often limit cross-school data sharing, which
hinders existing methods’ ability to model sparse data and address heterogeneity effectively, ultimately leading to suboptimal recommendations. In response, we propose HFRec, a heterogeneity-aware hybrid federated recommender system designed for cross-school elective course recommendations. The proposed model constructs heterogeneous graphs for each school, incorporating various interactions and historical behaviors between students to integrate context and content information. We design an attention mechanism to capture heterogeneity-aware
representations. Moreover, under a federated scheme, we train individual school-based models with adaptive learning settings to recommend tailored electives. Our HFRec model demonstrates its effectiveness in providing personalized elective recommendations while maintaining privacy, as it outperforms state-of-the-art
models on both open-source and real-world datasets. 
\end{abstract}


\section{Introduction}

In recent years, the education landscape has witnessed a rapid evolution, driven by advancements in technology and the increasing diversity of learners across institutions \cite{b1,b2,b3}. One critical aspect of modern education is the personalization of learning experiences, particularly in the context of elective course selection. Personalized course recommendations can significantly enhance students' learning outcomes by catering to individual preferences, abilities, and career aspirations. However, cross-school learner diversity creates unique challenges for recommender systems, as data sparsity and heterogeneity become significant obstacles to developing accurate and effective recommendations.

The exchange of information between schools could potentially alleviate these challenges. However, privacy concerns often inhibit data sharing, as sensitive information about students and their academic records must be protected. Existing methods, such as collaborative filtering \cite{b4}, content-based filtering \cite{b5}, and matrix factorization \cite{b6}, often struggle to model the sparse data and heterogeneous information available within individual schools. The limited data available for each school can lead to overfitting and cold-start problems, reducing the quality of recommendations \cite{b7}. Furthermore, these traditional approaches fail to capture the intricate relationships and diverse patterns prevalent in cross-school scenarios, ultimately resulting in suboptimal recommendations.

More recent developments, such as graph-based methods \cite{b8,b9,b10} and deep learning techniques \cite{b11}, have shown promise in handling sparse and heterogeneous data. However, these approaches often require substantial computational resources and still face challenges when applied to cross-school settings. For instance, they may not effectively model the diverse interactions and multifaceted relationships that arise due to the heterogeneity of cross-school data. Additionally, many of these methods assume that data can be centralized, which is not feasible in scenarios where privacy concerns restrict data sharing.

Federated learning \cite{b12} has emerged as a promising approach to address privacy concerns in collaborative learning settings. While federated learning has been successfully applied to various domains, its application to cross-school recommender systems remains under-explored. Moreover, existing federated learning-based recommender systems \cite{b13} often do not sufficiently model the heterogeneity inherent in cross-school data, limiting their performance in providing accurate and personalized elective recommendations.

To address these issues, there is a pressing need for a novel approach that can effectively handle data sparsity and heterogeneity while respecting privacy constraints. This approach should not only leverage the advantages of federated learning but also incorporate advanced techniques to model the intricate relationships and diverse patterns in cross-school settings.

In this paper, we propose HFRec, a heterogeneity-aware hybrid federated recommender system designed explicitly for cross-school elective course recommendations. Our approach circumvents the need for direct data sharing among schools, thereby addressing privacy concerns. The key contributions of this work can be summarized as follows:

\begin{itemize}
    \item We construct heterogeneous graphs for each school, which encompass various interactions and historical behaviors between students. These graphs facilitate the integration of context and content information, providing a rich and comprehensive representation of the underlying data.
    
    \item To capture heterogeneity-aware representations, an attention mechanism is designed to model diverse relationships between students and their interactions. It enhances recommendation accuracy while ensuring privacy through a federated approach and adaptive learning settings, enabling customized elective recommendations without direct data sharing among schools.
    
    \item Extensive experiments on open-source and real-world datasets validate the superiority of our HFRec model. It outperforms state-of-the-art models, providing personalized elective recommendations while maintaining privacy.
\end{itemize}
 
\section{Related Work}
\subsection{Graph Embedding for Recommendation}
Graph neural networks (GNNs) have gained popularity in recommendation systems for modeling complex user-item relationships. Various approaches have been proposed to leverage GNNs for learning user and item representations from user-item graphs, leading to more accurate recommendations. Approaches like GC-MC \cite{b14}, PinSage \cite{b15}, and NGCF \cite{b16} leverage GNNs to learn user and item representations from user-item or item-item graphs.

Other methods incorporate additional graph types, such as user-item-entity graphs \cite{b17} and user-user-item graphs \cite{b18}. KGAT \cite{b19} learns from a heterogeneous graph linking knowledge graph entities and user-item graph items, while GraphRec \cite{b18} captures social connections and preferences using user-item and user-user graphs.

Unlike these methods, which rely on centralized storage of user interactions and pose privacy concerns, our HFRec method ensures better privacy by keeping user data on local devices. This is crucial in educational settings with restricted data sharing between schools, safeguarding sensitive student information. Additionally, HFRec utilizes rich semantic information to recommend elective courses based on students' daily learning behaviors.

\subsection{Attention Mechanisms}
Attention mechanisms have achieved success in various tasks, including image captioning \cite{b20} and machine translation \cite{b21}. These mechanisms allow models to focus on relevant parts of the input, improving interpretability. Recently, attention has been incorporated into recommender systems \cite{b22,b23}, such as Attentional Factorization Machines (AFM) \cite{b6}, which identify the significance of feature interactions in content-aware recommendations.

In existing models, attention techniques are often used as supplementary components, such as attention+RNNs or attention+FMs. The Transformer model \cite{b24}, a well-known attention-based sequence-to-sequence method, has achieved state-of-the-art results in machine translation, surpassing traditional RNN/CNN-based approaches \cite{b25,b26}. Transformers heavily rely on self-attention modules to capture complex structures in sentences and determine relevant words for generating subsequent words in a sequence.

While there are models that combine Graph Neural Networks (GNNs) and Transformers \cite{b27,b28}, early Graph Transformers may underperform compared to Transformers when processing sequential data. Additionally, Graph Transformers require more memory and computational resources for training and inference due to their higher computational complexity.

\subsection{Federated Learning}
Federated learning is an innovative approach in machine learning that enables the development of intelligent models while preserving privacy \cite{b29,b30}. Unlike traditional methods, federated learning keeps user data on local devices \cite{b31}. Each device maintains a local model and computes updates using its local data. These updates are then sent to a central server, which coordinates the training process by aggregating the updates into a unified form and adjusting the global model. The updated global model is distributed back to the user devices to update their local models. This iterative process continues until convergence. By keeping raw user data local and sharing only model updates, federated learning significantly reduces privacy risks \cite{b32}.

Federated learning has also been applied to personalized recommendation systems. For example, FCF by Ammad et al. \cite{b33} computes gradients of user and item embeddings on user devices, with item embedding updates performed on a central server. FedMF by Chai et al. \cite{b36} protects item embeddings using homomorphic encryption techniques. Nevertheless, these methods do not consider high-order interactions between users and items, potentially limiting the accuracy of user and item representations. In contrast, our work introduces a novel hybrid federated learning approach that captures high-order interactions between students and items.

\section{Data Description}
Around 980 schools in Hong Kong, including secondary, primary, and special schools, use the Web-based School Administration and Management System (WebSAMS) for education purposes. However, the data collected from these schools exhibit inherent heterogeneity across several dimensions including academic performance, learning behaviors, extracurricular activities, technology usage, and teaching practices. These differences stem from diverse grading systems, curriculum structures, teaching methods, school cultures, resource availability and technology integration. Recognizing and addressing this heterogeneity is crucial for developing accurate and personalized recommender systems tailored to the unique characteristics and needs of students from different schools.

Five datasets from five different schools on WebSAMS are used in our experiments, namely dataset1($D_1$), dataset2($D_2$),...,dataset5($D_5$), respectively. Each dataset represents local web-based data from a respective school, with privacy restrictions prohibiting data sharing among schools. Focusing on students in Hong Kong, who complete three years of junior secondary education followed by three years of senior secondary education, we extracted six years of data to encompass all current students. Each student in the dataset is assigned a unique student ID, and their behavior data includes activity records, descriptions of various activities, and corresponding dates. The Extra-Curricular Activities (ECA) data provide an overview of the diverse student activities and participation records at schools throughout each academic year, which are considered as the description information for students' learning behavior. Table \ref{tab1} summarizes the main statistics of the five datasets.
 
\begin{table}[t]
\caption{Data Statistic of the Collected Real-world Dataset}
\begin{center}
\begin{tabular}{ccccc}
\toprule  % 顶部线
Schools&\#Students&\#Courses&\#ECA Records&\#ECA Categories\\ 
\midrule  % 中部线
$D_1$&662&450&4852&144\\
$D_2$&570&738&9564&472\\
$D_3$&1139&504&10233&144\\
$D_4$&1522&477&15265&323\\
$D_5$&785&463&2815&80\\
\bottomrule  % 底部线
\end{tabular}
\label{tab1}
\end{center}
\end{table}

\begin{table}[t]
\caption{Data Statistic of MOOC}
\begin{center}
\begin{tabular}{cc|cc}
\toprule  % 顶部线
Entities&Statistic&Relation&Statistic\\ 
\midrule  % 中部线
Student activities&48640&Student-course&319948\\
\midrule  % 中部线
Courses&706&Student-video&682753\\
\midrule  % 中部线
Videos&38181&Course-video&8951244\\
\bottomrule  % 底部线
\end{tabular}
\label{tab2}
\end{center}
\end{table}
We understand that it is important to evaluate our method on widely used and open-source datasets. However, due to the prevalent use of centralized databases in current online education platforms, there is lack of publicly available datasets for recommender systems utilizing federated learning. Therefore, We utilized the Massive Open Online Courses (MOOCs) dataset as a substitute for publicly available datasets in federated learning recommender systems. which can be downloaded from \href{http://moocdata.cn/data/MOOCCube}{http://moocdata.cn/data/MOOCCube}. Table \ref{tab2} shows the statistics of the data collected from MOOC. To preprocess the benchmark datasets for our experiments, we performed basic preprocessing on the MOOC dataset, creating five distinct datasets representing different schools. The size of each dataset was randomly selected within a specified range to mimic real-world scenarios. While the MOOC data lacks information on students' learning behavior during a course, we can consider the descriptive information related to students' video watching activities (e.g., video content, subtitles, viewing date and time) as content features similar to the data from Hong Kong schools.

To maximize student engagement in the recommended courses, we consider two metrics for evaluating students' ratings of the courses in MOOC and the real-world dataset, respectively. The first metric, course duration time, is measured by the ratio $r_{s,c} = \frac{t}{T}$, where $t$ represents the time a student spends on the course, and $T$ denotes the total course duration. The second metric, course score, is measured by the ratio $r_{s,c} = \frac{A_s}{A_c}$, where $A_s$ signifies the student's score in the course, and $A_c$ represents the average score of all students enrolled in that course. 

\section{Methodology}
In this section, we present a detailed explanation of our Hybrid Federated Recommender System (HFRec), designed for the purpose of recommending elective courses to individual students across various schools.

\subsection{Problem Formulation}\label{AA}
Given a target student with corresponding interactive data in our datasets, the goal is to calculate the rating about the student and the series of courses within the school. The recommend results is a top $N$ list of elective courses (e.g.,"Chinese history","English literature",...). 

Formally, denote $U, S, C$ as the sets of schools, students and courses respectively. Each school contains a set of students $S={s_1,,...,s_n}$. Besides, each school offers students a set of elective courses with descriptions $C={c_1,...,c_m}$ and a set of learning activities with descriptions $A={a_1,...,a_l}$.
Given the course enrollment history of each student and the corresponding learning performance $(s, c, p)$, where $s \in S$ and $c \in C$ and learning activity participating history of each students $(s,a)$, where $s \in S$ and $a \in A$. We assume there is not data sharing between schools and the full list of elective courses and learning activities are known by all schools. The objective is to learn a school-based function $f^{fed}(u)\rightarrow \{c_i|c_i\in C,i<K\}$ for any school $s\in S$, where $K$ is the length of the recommendation list consisting of candidate elective courses that students have not taken and can get good learning performance. 

\begin{figure*}[t]
\centerline{\includegraphics[width=\linewidth]{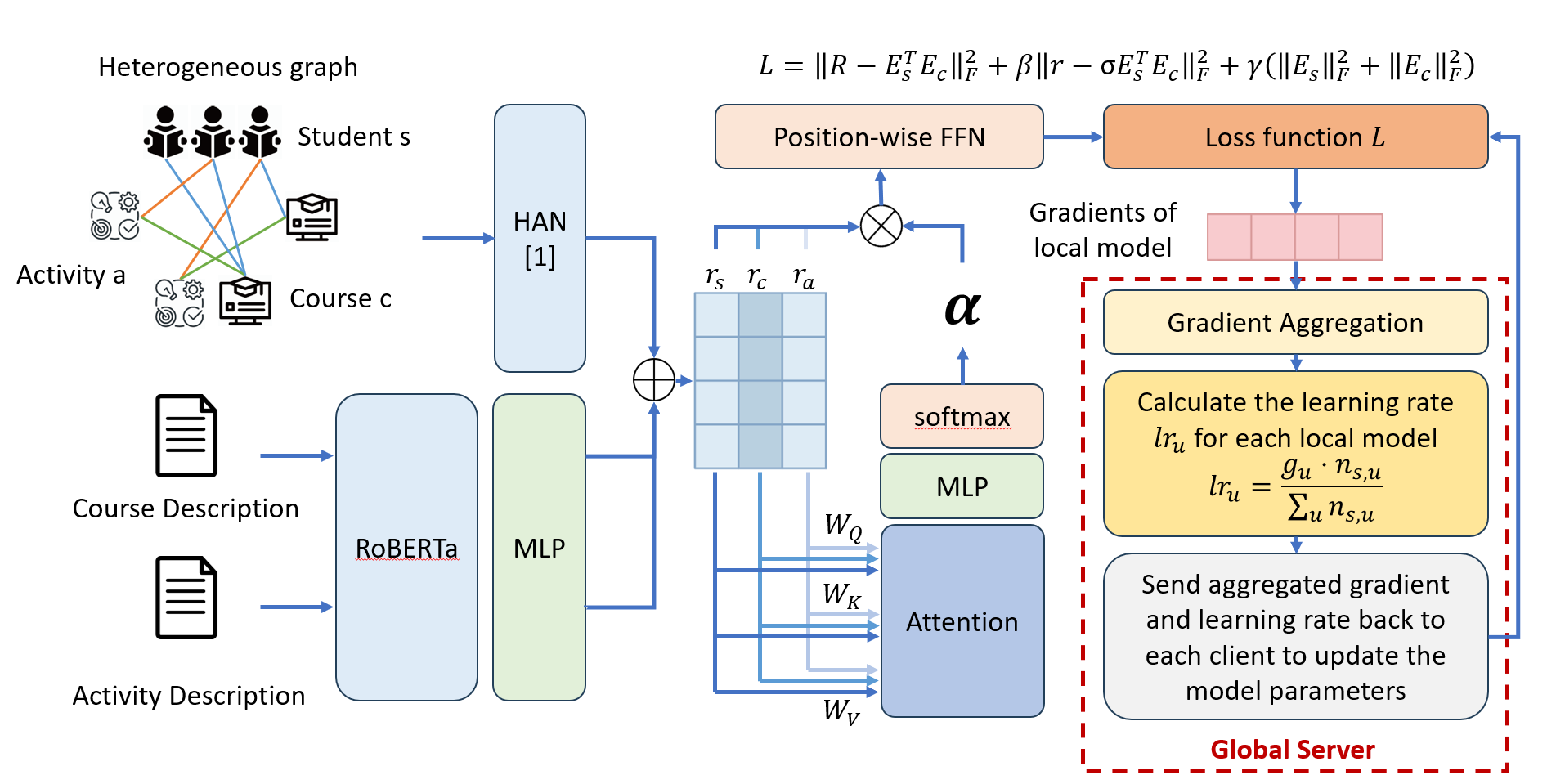}}
\caption{The system framework of HFRec .}
\label{fig1}
\centering
\end{figure*}

The architecture of our proposed hybrid federated recommender system, HFRec, is shown in Figure \ref{fig1}. The comprehensive model mainly consists of the following components:
\begin{itemize}
\item \textbf{Feature Extraction}: This component identifies relevant patterns and characteristics from raw data, reducing dataset complexity. It incorporates both content and context information to enhance the comprehension of the underlying structure and improve data representation.
\item \textbf{Representation Learning}: This component focuses on discovering meaningful representations of the extracted features. It utilizes a heterogeneity-aware attention mechanism for graph embedding, optimizing the downstream recommendation algorithm.
\item \textbf{Federated Model Update}: This component enables decentralized learning across schools, allowing collaborative model training and refinement while ensuring data privacy and minimizing centralized data storage. An adaptive learning rate balances varying dataset sizes, resulting in a personalized model for each school's students.
\end{itemize}

\subsection{Feature Extraction}
\textbf{Text Encoding:} To process the text descriptions of student activities written in traditional Chinese, we employed a RoBERTa-based model \cite{b34} for tokenization and conversion of token IDs to tensors. This allowed us to compute content-based embeddings for words and texts, which were then used as inputs for our representation learning models. For example, rich semantic information from names of courses(e.g., “Intensive Remedial Teaching Program,” “Applied Mathematics(ASL),” “Aesthetic Development”) and the descriptions of students' learning behaviors(e.g., "Participate in Robot Manufacturing Society“) are converted into vectors representations. For a given text $t$, its embedding 
\begin{equation}
    \textbf{e} = f_{RoBERTa}(t),\label{eq1}
\end{equation}
where $f_{RoBERTa}(t)$ denotes the maximum value of each feature across all token's hidden states in the last layer of the pre-trained RoBERTa \cite{b34}, therefore it can capture the most salient features of the input tokens.
Then the outputted embeddings are fed into a dense layer for dimension reduction:
\begin{equation}
    \overline{\textbf{E}} = RELU(\textbf{W} \textbf{E}_t)+\textbf{b},\label{eq2}
\end{equation}
where $RELU$(Rectified Linear Unit) is an activation function, $\textbf{W}$ is the weight matrix and $\textbf{b}$ is the bias vector.
Derived from \eqref{eq2}, the corresponding embeddings of the courses and learning behavior descriptions are: $\overline{\textbf{E}_c}$ and $\overline{\textbf{E}_l}$, respectively.\\ 

\textbf{Heterogeneous Information Modeling}: In addition to content features, latent contextual features can be leveraged by examining relationships among students, their learning behaviors, and courses. For example, students enrolling in the same course or participating in the same extracurricular activity may share common interests. To model these relationships, we construct a local heterogeneous graph for each school. This graph captures contextual information. Before discussing our approach, we will first introduce some relevant definitions.

\begin{itemize}
    \item A heterogeneous graph can be defined as a tuple: $\mathcal{G}=(V, R, T_V, T_R, \phi_V, \phi_R)$.
    \item $V$ is the set of nodes in the graph, $V = {v_1, v_2, ..., v_i}$.
    \item $R$ is the set of edges in the graph, $R \subseteq V \times V$. Each edge $r \in R$ is represented as an ordered pair $(v_i, v_j)$ where $v_i, v_j \in V$.
    \item $T_V$ is the set of node types, $T_V = \{t_{v,1}, t_{v,2}, ..., t_{v,k}\}$.
    \item $T_R$ is the set of edge types, $T_R = \{t_{r,1}, t_{r,2}, ..., t_{r,l}\}$.
    \item $\phi_V: v \rightarrow T_v$ is a function that maps each node $v \in V$ to its corresponding node type $t_v \in T_V$.
    \item $\phi_R: R \rightarrow T_R$ is a function that maps each edge $r \in R$ to its corresponding edge type $t_r \in T_R$.
\end{itemize}

Specifically, we consider the following different types of nodes and edges in a school's local graph $\mathcal{G}_s$:

\begin{itemize}
    \item Node $v_s$ : Each node $v_{s,i}$ represents a student $s_i$ within the school $u$.
    \item Node $v_c$: Each node $v_{c,i}$ represents a course $c_i$ that a student takes during school.
    \item Node $v_a$: Each node $v_{a,i}$ represents an activity with text describing a student 's learning behaviors (watching course video or participating an ECA).
    \item Edge $r_{u,c}$: The edge between node $v_s$ and node $v_c$ represents that relationship between a student $s$ and his/her interacted course $c$. 
    \item Edge $r_{s,a}$: The edge between node $v_s$ and node $v_a$ represents that relationship between a student $s$ and his/her interacted activity $a$. 
    \item Edge $r_{c,l}$: Lastly, the edge between node $v_c$ and node $v_a$ implies that a student take the course $c$ and participate activity $a$ at the same time. 
\end{itemize}

Based on the above definitions, we generate these relationships to provide more information for the node interaction and generate a fully connected heterogeneous graph to effectively model the sparse data from each local school. To capture the features of different types of nodes, an embedding layer $\textit{GE}$ is first used to convert the student node, course node and activity node into their embeddings $\textbf{E}_u$, $\textbf{E}_c$, $\textbf{E}_l$, respectively. 
\begin{equation}
     \textbf{E}_i = GE(\textbf{v}_i), i \in (u,c,l), \label{eq3}
\end{equation}
To fuse the contextual information into these node representations, we use element-wise addition to add the content feature and graph-based node embeddings element-wise: 
\begin{equation}
    \widehat{\textbf{E}} =  \overline{\textbf{E}} + \textbf{E},\label{eq4}
\end{equation}
where $\widehat{\textbf{E}}$, $\overline{\textbf{E}}$, $\textbf{E}$ denote the fused-embeddings, content-feature-embeddings and graph-based-embeddings, repectively. Noted that the content and context(graph-based) embeddings should have the same dimension which is pre-defined and we assume that the features are compatible and can be combined linearly. For the student nodes' embeddings, since they do not include any contextual information, their fused embeddings remain the same as the initial embeddings.

\subsection{Attention-based Representation Learning}
\subsubsection{Heterogeneity-aware Attention Mechanism} The adapting attention mechanism in the local heterogeneous graph within each school modifies the attention mechanism to handle different types of nodes and edges. It incorporates additional parameters for each node and edge type when calculating attention scores, capturing the relationships between elements in the graph.

The scaled dot-product attention mechanism \cite{b24} can be expressed as:
\begin{equation}
Attention(\textbf{Q}, \textbf{K}, \textbf{V}) = softmax(\frac{\textbf{QK}^T}{\sqrt{d}})V,\label{eq5}
\end{equation}
where $\textbf{Q}$ symbolizes queries, $\textbf{K}$ stands for keys, and $\textbf{V}$ denotes values, with each row corresponding to an individual item. The $\sqrt{d}$ scaling factor helps prevent inner product values from becoming excessively large. The self-attention (\textit{Att}) uses the embedding $\widehat{\textbf{E}}$ as input, transforms it into three distinct matrices via linear projections:

\begin{equation}
    S = Att(\widehat{\textbf{E}}\textbf{W}^Q, \widehat{\textbf{E}}\textbf{W}^K, \widehat{\textbf{E}}\textbf{W}^V),\label{eq6}
\end{equation}
where $\textbf{W}^Q,\textbf{W}^K,\textbf{W}^V \in \mathbb{R}^{d\times d}$ are the projection matrices. For each node in the graph, Query $(\textbf{Q})$, Key $(\textbf{K})$, and Value $(\textbf{V})$ vectors are generated by multiplying the input features with learned weight matrices $(\textbf{W}^Q, \textbf{W}^K, \textbf{W}^V)$. In our implementation, the compatibility scores between each pair of nodes $v_i$ and $v_j$ are calculated using their node embeddings $\widehat{\textbf{E}}_i$ and $\widehat{\textbf{E}}_j$ as well as the edge embeddings $\widehat{\textbf{E}}_{ij}$. Since we only want to learn the node representations of the students and courses, the corresponding edge embeddings can be considered as the node embeddings of $v_l$. Then a neural network is first used to take the concatenation of the node and edge embeddings as input:
\begin{equation}
    score(v_i, v_j) = neuralnet(concat(v_i, v_j, v_l))
\end{equation} \label{eq7}
This score captures the compatibility between nodes $v_i$ and $v_j$ considering both their features and the edge information. Secondly the Softmax function is applied to the compatibility scores to obtain the attention weights, ensuring that the attention weights for each node sum to 1:
\begin{equation}
    \alpha_{v_i,v_j} = Softmax(score(u, v))
\end{equation} \label{eq8}
Thirdly, multiply the attention weights by the value embeddings of the neighboring nodes:
\begin{equation}
    W_{v_i,v_j} = \alpha_{ij} \times \widehat{\textbf{E}}_{ij}
\end{equation} \label{eq9}
This step adjusts the influence of each neighboring node based on the computed attention weights, reflecting the importance of their features and the edge information. Lastly, sum the weighted values for all neighbors of each node $v$:
\begin{equation}
    \widehat{\textbf{E}}^{new}_{ij} = sum(W_{v_i,v_j})
\end{equation} \label{eq10}
This aggregation step combines the information from the neighboring nodes, incorporating both node features and edge information. The node $v$'s embedding is updated as $\widehat{\textbf{E}}^{'}_{ij}$.

\subsubsection{Position-wise feedforward layers} After the attention layer, position-wise feedforward layers are applied to the node embeddings. These layers consist of two linear transformations with an activation function (e.g., ReLU) in between.

\subsubsection{Layer normalization and residual connections} Our model also utilizes layer normalization and residual connections, similar to the original Transformer model \cite{b24}. Layer normalization is applied to the node embeddings before and after the multi-head attention and position-wise feedforward layers. Additionally, residual connections are used to combine the input and output of these layers. The output of the layers would be a set of students' vectors: $\textbf{E}_{s}$ and a set of courses' vectors: $\textbf{E}_{c}$.

\subsection{Federated Learning Scheme}

Federated learning is a decentralized approach that enables multiple clients to train a shared model while keeping their data private and local. This protects data privacy, reduces the risk of data breaches, and promotes collaboration without sharing sensitive information. Our proposed federated learning strategy for the recommender system considers data heterogeneity and consists of three main components.

\subsubsection{Constrained Matrix Factorization for Recommendation}
Matrix Factorization (\textbf{MF}) is a popular method in recommender systems, where a matrix \textit{R} is factorized into two lower-dimensional matrices \textit{U} and \textit{V}. These matrices are used to fill in missing values in the original matrix. However, the uneven distribution of datasets among schools can lead to faster training for elective courses with higher selection rates (\textbf{SR}), while courses with low \textbf{SR} are not fully trained, and increasing training times could lead to overfitting for courses with high \textbf{SR}. To address this, we utilize a Constrained Matrix Factorization (\textbf{ConMF}) technique \cite{b35}, which enhances traditional Matrix Factorization by considering the course average rating. Specifically, an error term is incorporated into the objective function to constrain the latent factor matrices, allowing predicted scores to approach the course average scores and reducing the impact of course selection rates (\textbf{SR}). Our goal is to minimize the difference between predicted and real ratings while considering variations in data across schools. The proposed loss function is defined as follows:
\begin{equation}
    L = \mathop{\arg\min}_{\textbf{E}_{s},\textbf{E}_{c}} \| (R-\textbf{E}_{s}^T \textbf{E}_{c}) \|^2_F + \beta\|r-\sigma \textbf{E}_{s}^T \textbf{E}_{c}\|^2_F + \gamma \Omega 
\end{equation},\label{eq11}
where $R$ is the real rating matrix, $r \in \mathbb{R}^{1\times n}$ is a course average rating vector with the $k$-th entry in $r$ being the average rating of the $k$-th course. $\sigma \in \mathbb{R}^{1\times m}$ is a transfer vector and every element of $\sigma$ is equal to $\frac{1}{m}$ , where $m$ is the number of students. $\beta$ and $\gamma$ are two regularization parameters. $\Omega$ is the regularization term which is defined as:
\begin{equation}
    \Omega = \|\textbf{E}_{s}\|^2_F + \|\textbf{E}_{c}\|^2_F
\end{equation}\label{eq12}
Stochastic gradient descent is employed to update the two latent factor matrices $\textbf{E}_{s}$ and $\textbf{E}_{c}$. Then the predicted rating of student $s$ in course $c$ $P(s, c)$ would be:
\begin{equation}
    P = \widehat{\textbf{E}}_{s} \cdot \widehat{\textbf{E}}_{c} 
\end{equation}\label{eq13}

\subsubsection{Federated Learning Approach}
Building upon the Constrained Matrix Factorization (ConMF) approach, our objective is to minimize the loss for each local model and achieve accurate rating predictions for every local recommender system. To accomplish this, we introduce a coordinating server that oversees all participating schools and computes global gradients for updating the local models and embedding parameters.

In each iteration, the server selects a certain number of schools to calculate gradients locally and transmits them to the server. Upon receiving these local gradients, the server aggregates them into a unified gradient denoted as $g_u$. The aggregated gradient is then sent back to each client for local parameter updates. The parameter set for the $i$ th school $u_i$, denoted as $\Theta$, is updated using the formula $\Theta_i = \Theta_i - lr \times g_u$, where $lr$ represents the learning rate. This iterative process continues until the model converges. The structure of our proposed method, HFRec, is outlined in Algorithm 1

\begin{algorithm}[t]
	\renewcommand{\algorithmicrequire}{\textbf{Input:}}
	\renewcommand{\algorithmicensure}{\textbf{Output:}}
	\caption{HFRec}
	\label{alg:1}
	\begin{algorithmic}[1]
		\REQUIRE local graph for each school: $\mathcal{G}_i$ with its adjancy matrix: $\mathcal{A}_i$ 
        Initialize the graph embedding: $\Theta_i$ for each school
        \\//\textbf{Global}\\
        \textbf{Repeat}
		\STATE Select a subset $\overline{U}$ from school set $U$ randomly  
		\STATE $g$ = 0
		\FOR{each $u_i \in \overline{U}$}
		\STATE $g \leftarrow g + Local(\mathcal{G}_i)$
		\ENDFOR		
        \STATE Distribute $g$ to local model for local update\\
        \textbf{Until} model convergence
        \\//\textbf{Local}
        \FOR{each embedding $\widehat{E}$}
        \STATE Calculate $\widetilde{\Theta_i}$ via Graph embedding \eqref{eq4} and Attention layer\eqref{eq7}
        \ENDFOR
        \STATE Compute the model gradient $Local(\mathcal{G}_i)$ by minimizing $Loss(\widetilde{\Theta_i})$
		\STATE \textbf{return} $Local(\mathcal{G}_i)$
	\end{algorithmic}  
\end{algorithm}

\subsubsection{Adaptive Learning Rate}
In federated learning, clients possess varying amounts and distributions of data, which can result in uneven progress during local updates and hinder global model convergence. To address this heterogeneity, we adopt adaptive learning rates or local learning rate schedules that consider the data distribution of each client. By adjusting the learning rate based on the characteristics of each client's data distribution, we can effectively account for the variations and ensure better model learning from diverse data sources.

In our system, we scale the learning rate for each school based on the size of their local dataset. This can be done as follows: Firstly we calculate the global learning rate $lr_{global}$ for the federated learning process. Then, for each school $u$, calculate the local learning rate $lr_u  = lr_{global} \times \frac{n_u}{N}$, where $n_u$ is the number of samples in the local dataset of school $u$, and $N$ is the total number of samples across all schools. Lastly we use the calculated local learning rate $lr_u$ for each school during the local optimization process.

This approach ensures that schools with larger datasets contribute more to the global model update, as their learning rates are higher. Schools with smaller datasets will have lower learning rates, reducing the risk of overfitting to their local data.

\begin{table*}[t]
\caption{Comparison Results on MOOC}
\begin{center}
\begin{tabular}{cccccccccc}
\toprule  % 顶部线
MOOC&HR@1&HR@5&HR@10&HR@20&NDCG@5&NDCG@10&NDCG@20&MRR&AUC\\ 
\midrule  % 中部线
\#FCF&0.278&0.361&0.571&0.767&0.352&0.449&0.438&0.342&0.619\\
\midrule  % 中部线
\#FedMF&0.283&0.378&0.572&0.770&0.377&0.458&0.503&0.366&0.618\\
\midrule  % 中部线
\#FedGNN&0.345&0.433&0.613&0.838&0.499&0.502&0.554&0.457&0.745\\
\midrule  % 中部线
\textbf{\#HFRec}&\textbf{0.363}&\textbf{0.486}&\textbf{0.757}&\textbf{0.911}&\textbf{0.540}&\textbf{0.580}&\textbf{0.624}&\textbf{0.537}&\textbf{0.815}\\
\bottomrule  % 底部线
\end{tabular}
\label{tab3}
\end{center}
\end{table*}

\begin{table*}[t]
\caption{Comparison Results on DMP}
\begin{center}
\begin{tabular}{cccccccccc}
\toprule  % 顶部线
DMP&HR@1&HR@5&HR@10&HR@20&NDCG@5&NDCG@10&NDCG@20&MRR&AUC\\ 
\midrule  % 中部线
\#FCF&0.264&0.359&0.589&0.749&0.349&0.449&0.463&0.338&0.659\\
\midrule  % 中部线
\#FedMF&0.223&0.376&0.587&0.738&0.356&0.458&0.449&0.325&0.671\\
\midrule  % 中部线
\#FedGNN&0.267&0.426&0.679&0.821&0.431&0.502&0.621&0.518&0.789\\
\midrule  % 中部线
\textbf{\#HFRec}&\textbf{0.323}&\textbf{0.606}&\textbf{0.787}&\textbf{0.881}&\textbf{0.579}&\textbf{0.653}&\textbf{0.681}&\textbf{0.611}&\textbf{0.841}\\
\bottomrule  % 底部线
\end{tabular}
\label{tab4}
\end{center}
\end{table*}

\begin{figure*}[htbp]
\centerline{\includegraphics[width=\linewidth]{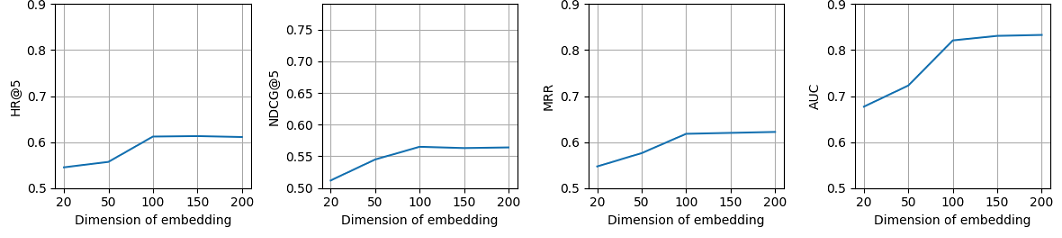}}
\caption{Performance of different dimension of embedding.}
\label{fig2}
\end{figure*}

\begin{figure*}[htbp]
\centerline{\includegraphics[width=\linewidth]{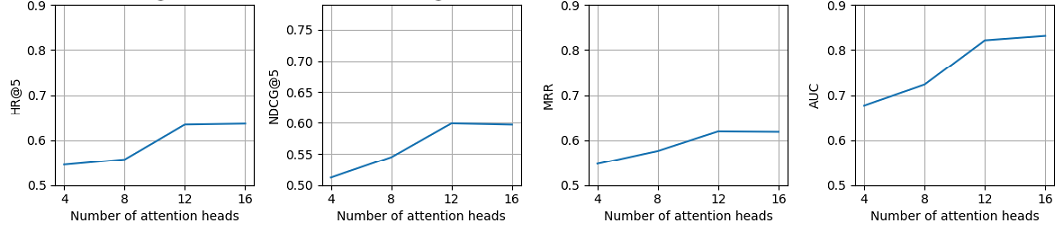}}
\caption{Performance of different number of attention heads.}
\label{fig3}
\end{figure*}
\section{Experiments}
In our experimental setup, we utilized datasets collected from MOOCs and WebSAMS, as explained in the previous section. The data from 2016 to 2020 was used as the training set, while the data from 2020 to 2022 served as the testing set. Each instance in the training and testing sets represents a sequence of a student's past activity behaviors.

During the training process, we considered the highest-rated courses selected by a student in each sequence of the training data as the target, while the remaining courses were treated as the student's past behaviors. For positive instances, we randomly generated negative instances by replacing the target course.

In the testing process, each enrolled course in the test set was considered as the target course. The corresponding courses of the same student in the training set represented the sequence of the student's historical course selections. To evaluate the recommendation performance, we paired each positive instance in the test set with 99 randomly sampled negative instances. The model then provided prediction scores for these 100 instances (1 positive and 99 negatives).

\subsection{Experimental Settings}
In our experiments, we use dot product to implement the rating
predictor. The user and item embeddings and their hidden representations learned by graph neural networks are 100-dim. The number of the attention heads $h$ is 12 and the total number of epoch \emph{e} is 3. The ratio of dropout is 0.2. SGD is selected as the optimization algorithm. The hyperparameters are selected according to the validation performance. 

Our experimental datasets retain the validation and test sets of the original dataset. We train our model on datasets collected from both Hong Kong and MOOC platforms, comparing the performance with state-of-the-art models. This serves as a preliminary to our experimental settings.

\subsection{Evaluation Metrics and Baseline Methods}
We evaluate all the methods in terms of the widely used metrics,
including Hit Ratio of top-K items $(HR@K)$ and Normalized Discounted Cumulative Gain of top-K items $(NDCG@K)$ \cite{b39}. $HR@K$ is a recall-based metric that measures the percentage of ground truth instances that are successfully recommended in the top K, and $NDCG@K$ is a precision-based metric that accounts for the predicted position of the ground truth instance. We set K to 5, 10, and 20, and calculate all metrics for every 100 instances (1 positive plus 99 negatives). The larger the value of $HR@K$ and $NDCG@K$, the better the performance of the model.
We also use the mean reciprocal rank ($MRR$). From the definition \cite{b40}, we can see that a larger $MRR$ value indicates a better performance of the model.
In addition, we also add the area under the curve of $ROC(AUC)$ as a metric.

\subsection{Performance Evaluation}
We compare the performance or our HFRec model with several privacy-reserving methods based on federated learning, including:
\begin{itemize}
    \item FCF \cite{b33}, a privacy-preserving recommendation approach
based on federated collaborative filtering;
    \item FedMF \cite{b36}, another privacy-preserving recommendation approach based on secure matrix factorization.
    \item FedGNN \cite{b37} a GNN-based federated learning approach.
\end{itemize}

Table \ref{tab3} and Table \ref{tab4} display the comparison results on the MOOC and DMP datasets, respectively. Our HFRec approach outperforms all other methods in all metrics on both datasets, demonstrating its ability to preserve user privacy while providing excellent recommendations.

HFRec leverages contextual information in a heterogeneous manner, effectively capturing diverse relationships and surpassing methods relying solely on content or context features. The adaptive fusion of representations, combining rich content features with structural relations between entities, contributes to HFRec's superior performance. Overall, the results confirm the effectiveness of our proposed method in capturing comprehensive information and providing accurate recommendations while preserving user privacy.

\subsection{Hyperparameter Analysis}
In this section, we present a hyperparameter analysis of our HFRec model, focusing on two key aspects: the dimension of the node embeddings and the number of attention heads used in the attention layer. First, we examined the effect of varying the dimension of embeddings between 50 and 200. The results in Figure \ref{fig2} indicate that the model's performance is sensitive to the choice of embedding size. The best performance was achieved with an embedding dimension of around 100, suggesting that this value provides a suitable balance between model complexity and the ability to capture meaningful patterns in the data.

Next, we analyzed the impact of varying the number of attention heads, ranging from 4 to 16. The attention mechanism is a crucial component of the HFRec model, enabling it to efficiently capture various types of relationships in the data. Analysis results in Figure \ref{fig3} reveal that the optimal number of attention heads is around 12, as increasing the number beyond this point did not yield significant performance improvements. This finding suggests that 12 attention heads provide sufficient capacity for the model to capture relevant information, while keeping the computational cost manageable.

\subsection{Case Study}
In this part, we conduct one case to demonstrate the effectiveness of our proposed method HFRec. We randomly select a $student:4188$ from school $s_1$ and obtain a top 5 recommend lists of courses.

\begin{figure}[t]
\centerline{\includegraphics[width=1\linewidth]{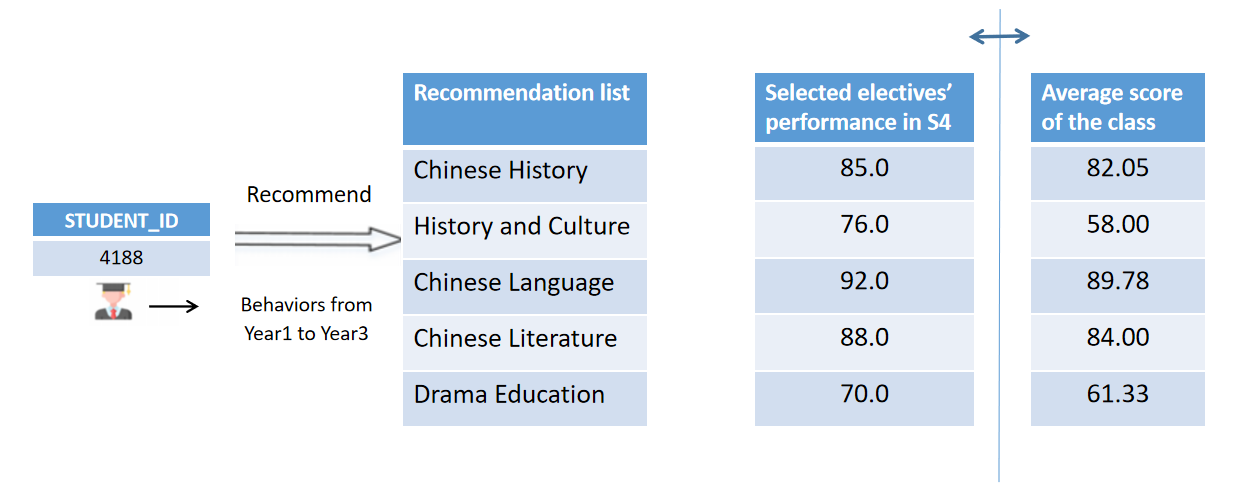}}
\caption{The case study of HFRec of a student.}
\label{fig4}
\end{figure}

We employed the student's historical behaviors from year 1 to 3 as input for our recommender system, which generated a list of five elective courses tailored to the student's interests. Upon examining the student's academic performance in these recommended courses during year 4, we observed that the student performed exceptionally well. Figure \ref{fig4} demonstrates the observation of the student's academic performance, which indicates that our recommendation strategy effectively engaged the student and aligned with their interests. The results demonstrate that our privacy-preserving recommender system successfully captures students' interests and makes a significant contribution to the field of education.

\section{Conclusions}
In conclusion, HFRec is a heterogeneity-aware hybrid federated recommender system that accurately captures students' preferences for cross-school elective course recommendation. It incorporates content-based features, contextual information, and a heterogeneity-aware attention mechanism. Utilizing federated learning enhances data security and enables distributed training across schools. This work builds upon prior research in hybrid recommender systems, graph-based methods, and federated learning, providing an interpretable and privacy-preserving solution. Despite employing federated learning to address data privacy and sparsity issues, privacy risks may still exist. Although our approach avoids direct sharing of training data between schools, there is a possibility of sensitive information being present in the exchanged data. Future work can explore more efficient federated learning algorithms and conducting experiments to evaluate the model's privacy preservation capabilities would provide valuable insights.

\section*{Acknowledgment}
This research was conducted in Research Institute for Artificial Intelligence of Things (RIAIoT) at the Hong Kong Polytechnic University (PolyU) and was supported by Hong Kong Jockey Club Charities Trust (Project S/N Ref.: 2021-0369) and PolyU Research and Innovation Office (No. BD4A).

\end{document}